\documentclass[aps,prb,twocolumn,showpacs,superscriptaddress,groupedaddress]{revtex4}  
\usepackage{graphicx}  
\usepackage{dcolumn}   
\usepackage{bm}        
\usepackage{amssymb}   
\usepackage{color}
\usepackage{pdfpages}

\begin{document}

\title{One-way Tamm plasmon-polaritons on the interface of magnetophotonic crystals and conducting metal oxides}

\author{Hui Yuan Dong} \affiliation{Department of Physics, Southeast University, Nanjing 211189, China}\affiliation{School of Science, Nanjing University of Posts and Telecommunications, Nanjing 210003, China}
\author{Jin Wang} \affiliation{Department of Physics, Southeast University, Nanjing 211189, China}
\author{Tie Jun Cui} \email{tjcui@seu.edu.cn} \affiliation {State Key Laboratory of Millimeter Waves, Department of Radio Engineering, Southeast University, Nanjing 210096, China}

\date{\today}

\begin{abstract}

We demonstrate theoretically the existence of one-way Tamm plasmon-polaritons on the interface between magnetophotonic crystals and conducting metal oxides. In contrast to conventional surface plasmon-polaritons (SPPs), Tamm plasmon-polariton (TPPs) occur at frequencies above the bulk plasma frequency of the conducting materials, provided that the dispersion curves of such surface modes lie outside the light cone for the conducting oxides and simultaneously fall into the photonic band gap of the magnetophotonic crystal. The nonreciprocal properties of TPPs are caused by violation of the periodicity and time reversal symmetry in the structure. Calculations on the field distribution and transmission spectra through the structure are employed to confirm the theoretical results, which could potentially impact on a broad range of SPP-related phenomena in applications.

\end{abstract}

\maketitle

\section{\label{sec:level1}INTRODUCTION}

There have been strong interests in the surface state predicted by Tamm in 1932 \cite{Tamm}, due to its key role in understanding various fundamental properties of solids. In analogy with purely electronic states in a semiconductor, optical surface states which can occur on the interface of optical superlattices were later analyzed theoretically and demonstrated experimentally by Yeh $et ~al.$ \cite{Yeh1,Yeh2}. Such optical states were then called as optical Tamm states (OTSs) \cite{Vinogradov} to distinguish from the purely electronic excitation. The first experimental verification of OTSs on the interface of magnetophotonic and nonmagnetic photonic crystals was reported by Goto $et ~al.$ in 2008 \cite{Goto}.

Surface plasmon-polaritons (SPPs) are a related surface state formed on the boundary of metallic and dielectric media \cite{Kawata}. The confinement in the metal is the result of the metal's negative dielectric constant at frequencies below its bulk plasma frequency, while the confinement in the dielectric media is due to the total internal reflection. These surface plasmons propagate along the metal surface with frequencies ranging from zero towards the asymptotic value $\omega_{p}/\sqrt{2}$, where $\omega_{p}$ is the bulk plasma frequency. The dispersion curve of the surface plasmon lies to the right of the light line (given by $\omega=c k_x$, where $k_x$ is the in-plane component of the wave vector of light and $\omega$ is the angular frequency), and therefore excitation by light beams is not possible unless special techniques for phase-matching, such as prism and grating coupling, are employed. Recently, another form of surface state, called Tamm plasmon polaritons (TPPs), was theoretically proposed  and experimentally confirmed by Kaliteevski $et ~al.$ \cite{Kaliteevski,Shelykh,Sasin}. In contrast to a conventional surface plasmon polariton, Tamm plasmon polaritons can be formed in both the TE and TM polarization, and be observed at the interface between a metal and dielectric Bragg mirror without the use of prism, grating coupling or the alternative surface structuring approach. Moreover, the confinement in the dielectric multilayer structure is due to the photonic band gap of the Bragg mirror, instead of total internal reflection. Potential applications on TPPs \cite{Gong,Zhou,Zhang,Symonds,Grossmann,Liew,Gazzano,Kavokin} have been found in the realization of optical components, such as absorbers \cite{Gong}, filters \cite{Zhou}, and bistable switches \cite{Zhang}. It has also been found that one can achieve strong coupling between TPPs and the excitons from a quantum well \cite{Symonds, Grossmann} or a quantum dot \cite{Gazzano}, and possibly utilize these states in the construction of a polariton laser without a microcavity \cite{Kavokin}.

On the other hand, SPPs may exhibit nonreciprocal properties in the presence of an external magnetic field. For example, Yu $et ~al.$ \cite{zongfu} demonstrated the existence of one-way electromagnetic waveguides formed at the interface between a plasmonic metal under a static magnetic field and a photonic crystal. Such a waveguide provides a frequency range where only one propagating direction is allowed. However, the working frequencies are limited to be lower than the bulk plasma frequency of metal, satisfying the conditions that SPPs are bounded at the interface, that is the permittivity of the metal should be negative.

In surface-plasmon studies, gold and silver are typically employed with plasma frequencies above the visible or infrared part of electromagnetic spectrum. Recent experimental work has revealed the surface-plasmon resonance phenomenon could be observable in a conducting metal oxides thin film, with the smaller bulk plasma frequency (i.e. $\hbar\omega_{p}=1$ eV for indium tin oxide). In this work, we aim to discuss the nonreciprocal properties for TPPs that may exist at an interface of magnetophotonic crystals (MPCs) and conducting metal oxides, at frequencies above the bulk plasma frequency of the conducting materials. In this case, although the permittivity of conducting oxides is positive in the working frequencies, TPPs can be bounded at the interface of conducting oxides by total internal reflection on one side, and by the photonic band gap (PBG) of MPCs on the other. Furthermore, the dispersion curves of TPPs modes lie in part inside the light cone for free space, and such TPPs can be excited under direct illumination of a plane wave. We further calculate the field pattern and transmission spectra through the structure to support the spectral splitting in the dispersion of wave propagating in the opposite directions.

\section{\label{sec:level1}MODEL AND METHODS}

Let us begin for the structure shown in Fig. 1, with a semi-infinite conducting region on the left ($z<0$) and a two-component MPCs on the right ($z>0$). The unit cell of MPCs consists of one isotropic dielectric and one magnetooptical layer, with thickness $l_a$ and $l_b$, respectively. The relative permittivity of the conducting material is taken to be of the Drude form
\begin{equation}
\epsilon_{c}(\omega)=1-\frac{\omega_{p}^{2}}{\omega(\omega+i\gamma)},
\end{equation}
where $\gamma$ is the electronic collision frequency. For initial calculations, we take $\gamma=0$, but subsequently we consider the effect of losses by assuming $\gamma\neq0$. For the bulk plasma frequency of the conducting material, we just choose $\hbar\omega_{p}=1$ eV \cite{Brand}, which can occur in materials such as indium tin oxide.

\begin{figure}[!htbp]
\includegraphics[width=4in]{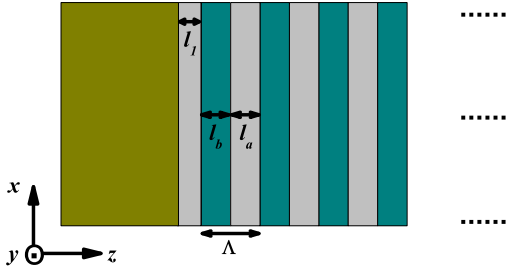}
\caption {\small (color online) Schematic diagram of the structure, comprising a semi-infinite metallic region on the left and a semi-infinite MPCs on the right with their interface at $z=0$. The period of MPCs is $\Lambda$.}
\end{figure}

To accomplish the required symmetry breaking, we use a gyrotropic material, Bismuth iron garnet (BIG) for magnetooptical layer in the MPCs. The optical property of BIG is characterized by a dielectric tensor
\begin{equation}
\bar{\epsilon}_{b}=
\left(
\begin{tabular}{ccc}
$\epsilon_{b}$ & $0$ & $i\Delta_{b}$\\
$0$ & $\epsilon_{b}$ & $0$\\
$-i\Delta_{b}$ & $0$ & $\epsilon_{b}$\\
\end{tabular}
\right),
\end{equation}
when the magnetization is along the $y$ direction. As a consequence, TM and TE modes ($H$ or $E$ fields polarized along the $y$ direction, respectively) are completely decoupled. Therefore, the one-way TPPs modes will keep the TM behavior in this configuration. The use of isotropic dielectric layer, such as SiO$_2$ glass (also characterized by the dielectric and magneto-optical parameters $\epsilon_{a}$ and $\Delta_{a}$, respectively), provides good index contrast with BIG to create the band gap.

To show the one-way TPPs property, we find the dispersion of TPPs by using the standard transfer matrix approach \cite{Khanikaev,Liscidini}. First, we consider an infinite periodic structure of MPCs, the transfer matrix associated with it is
\begin{equation}
\hat{T}=
\left (
\begin{tabular}{cc}
$T_{11}$ & $T_{12}$ \\
$T^{*}_{12}$ & $T^{*}_{11}$ \\
\end{tabular}
\right)
=
\hat{P}_a \hat{M}_{ba} \hat{P}_b \hat{M}_{ab}.
\end{equation}
The $\hat{P}_{i}$ are the usual propagation matrices,
\begin{equation}
\hat{P}_{i}=
\left (
\begin{tabular}{cc}
$e^{i k_{zi} l_{i}}$ & $0$ \\
$0$ & $e^{-i k_{zi} l_{i}}$ \\
\end{tabular}
\right),
\end{equation}
where
\begin{equation}
k_{zi}=\sqrt{(\frac{2\pi}{\lambda_0} n_{i})^2-k_{x}^{2}},
\end{equation}
with $\lambda_0$ the wavelength in vacuum, $k_x$ the component of the wave vector in the plane of surface, and $n_{i}=
\sqrt{(\epsilon_{i}^{2}-\Delta_{i}^{2})/\epsilon_{i}}$ the refractive index. The $\hat{M}_{ij}$ are the interface matrices,
\begin{equation}
\hat{M}_{ij}=\frac{\epsilon_{j}^{2}-\Delta_{j}^{2}}{2\epsilon_{j}k_{zj}}
\left (
\begin{tabular}{cc}
$F^*_{j}+F_{i}$ & $F^*_{j}-F^*_{i}$ \\
$F_{j}-F_{i}$ & $F_{j}+F^*_{i}$ \\
\end{tabular}
\right),
\end{equation}
where $F_{m}=(\epsilon_{m}k_{zm}+i\Delta_{m}k_x)/(\epsilon_{m}^{2}-\Delta_{m}^{2})$, $m=i,j$. The eigenvectors of $\hat{T}$ satisfy the relation
\begin{equation}
\hat{T}
\left (
\begin{tabular}{cc}
$a_0$ \\
$b_0$ \\
\end{tabular}
\right)
=e^{iK\Lambda}
\left (
\begin{tabular}{cc}
$a_0$ \\
$b_0$ \\
\end{tabular}
\right)
\end{equation}
where $\Lambda=l_a+l_b$ is the MPCs period and $K$ is the Bloch wave vector. We can take $a_0={T}_{12}$, and $b_0=e^{iK\Lambda}-{T}_{11}$.

More generally, we can take a unit cell in which a first layer of index $n_a$ with thickness $l_1=\sigma l_a$, where $\sigma\in[0,1]$. The general transfer matrix \cite{Liscidini} is given by $\hat{T}_{\sigma}=\hat{P}_{\sigma}^{-1}\hat{T}\hat{P}_{\sigma}$, where $\hat{P}_{\sigma}=$ diag$(e^{ik_{za}l_1},e^{-ik_{za}l_1})$. We can easily find that $\hat{T}$ and $\hat{T}_{\sigma}$ have the same eigenvalues. The solution of such eigenvalue problems gives Bloch modes of an infinite MPCs in an explicit form.

The TPPs dispersion relation is then obtained by the modal matching at the interface between the terminating layer of MPCs and the conducting metal oxides. Calculation gives the following dispersion for the TPPs \cite{Khanikaev, Liscidini}:
\begin{equation}
q_{c}=ik_{za}\frac{n_{c}^{2}}{n_{a}^{2}}\frac{T_{12}e^{-2ik_{za}l_1}+T_{11}-e^{iK\Lambda}}{T_{12}e^{-2ik_{za}l_1}-T_{11}+e^{iK\Lambda}}, \label{modes}
\end{equation}
where $q_{c}$ is defined as
\begin{equation}
q_{c}\equiv-ik_{zc}=\sqrt{k_{x}^{2}-(\frac{2\pi}{\lambda_{0}}n_{c})^2}
\end{equation}
with $n_c$ refractive index of conducting metal oxides.

\section{\label{sec:level1}Results}
To demonstrate the nonreciprocity of TPPs, we solve the Eq. (\ref{modes}) numerically for a specific semi-infinite MPCs with alternate layer of SiO$_2$ ($\epsilon_{a}=2.07$ and $\Delta_{a}=0$) and BIG ($\epsilon_{b}=6.25$ and $\Delta_{b}=0.06$), a period of $\Lambda=187$ nm. We take the parameter $\sigma=0.4$, which determines truncation of the terminating layer of the structure (here is SiO$_2$). For the sake of illustration we have used $\Delta_{b}=0.6$, which is ten times greater than the realistic material constant \cite{zongfu2}.

\begin{figure}[!htbp]
\centering
\includegraphics[width=2.2in]{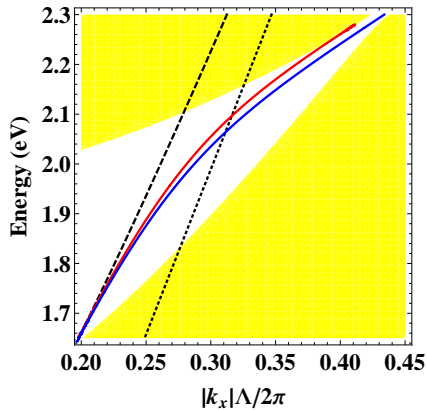}
\caption {\small (color online) Dispersion of TPPs at the interface of conducting metal oxides and semi-infinite MPCs. Red and blue lines correspond to the forward and backward propagating TPPs, respectively. Yellow and white regions correspond to bands and gaps of an infinite MPCs. Light curves for conducting metal oxides (dashed line) and free space (dotted line) are also shown.}
\end{figure}

We show in Fig. 2 the dispersion of the forward ($k_{x}>0$) and backward ($k_{x}<0$) TPPs in the first photonic bandgap by red and blue lines, respectively. The key result is that there exist asymmetric TPPs solutions, $\omega(k_x)\neq\omega(-k_x)$, which lie above the bulk plasma frequency of the metal (here given by $\hbar\omega_{p}=1$ eV). The spectral splitting of the dispersion of waves propagating in the opposite direction then gives rise to the nonreciprocal TPPs. Physically, such reciprocity develops from the magnetization as well as the violation of the periodicity in MPCs, which is directly related to the matrix elements $T_{11}$ and $T_{12}$ of the transfer matrix and Bloch wave vector $K$ in Eq. (\ref{modes}). For the range of result shown, the dispersion curves lie outside the light line for conducting oxides, and in part within the light line for free space, indicating the associated modes are bounded at the surface of conducting metal oxides, and also accessible to direct excitation by incident radiation without the need for prism or grating coupling.

\begin{figure}[!htbp]
\centering
\includegraphics[width=3in]{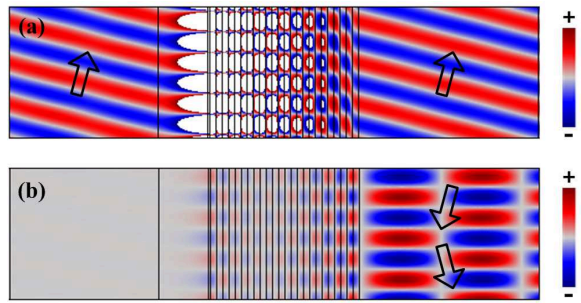}
\caption {\small (color online) Out-of-plane magnetic field patterns for the finite-size structure consisting of a conducting layer on the surface of a $12$ period MPCs, at the energy $E_{+}=2.054$ eV under front illumination (a) and back illumination (b), when the incident angle is taken to be $75.60^\circ$.}
\end{figure}

In order to give one a simple and instructive explanation of the discovered one-way character of TPPs, we plot the out-of-plane magnetic field profile in Fig. 3 for a finite structure consisting of a conducting layer on the surface of a $12$ period MPCs, under the light illumination of a plane wave. As an example, we respectively take the in-plane wave vector $k_{x+}=0.3* 2\pi/\Lambda$ for forward illumination, and $k_{x-}=-0.3* 2\pi/\Lambda$ for backward illumination, within the light line of free space for energies considered. For both of such in-plane wave vectors, the PBG of the infinite MPCs in Fig. 2 is between $1.899$ and $2.128$ eV for the TM polarization. The corresponding energies for TPPs at $k_{x+}$, $k_{x-}$ are $E_{+}=2.054$ and $E_{-}=2.034$ eV, respectively, and we thus expect to see the wave perfectly propagating through the structure near the energy $E_{+}$ when the light is incident from front at an angle $75.60^\circ$, or near $E_{-}$ when light from back at an incident angle, $77.97^\circ$.

The steady-state field patterns at the energy $E_{+}$ are shown in Fig. 3. Counter-propagating plane waves are incident from air upon either end of the MPCs. For the case of forward incidence, the field amplitude is remarkably high at the interface between conducting metal oxides and MPCs, and falls exponentially away from the interface. Such a distribution confirms the formation of the TPPs, providing complete transparency of the structure seen in Fig. 3(a). In contrast, for backward incidence in Fig. 3(b), such excitation of TPPs is almost completely suppressed, resulting in low transmission through the structure. Thus such a structure demonstrates the one-way total transmission.

\begin{figure}[!htbp]
\centering
\includegraphics[width=3.2in]{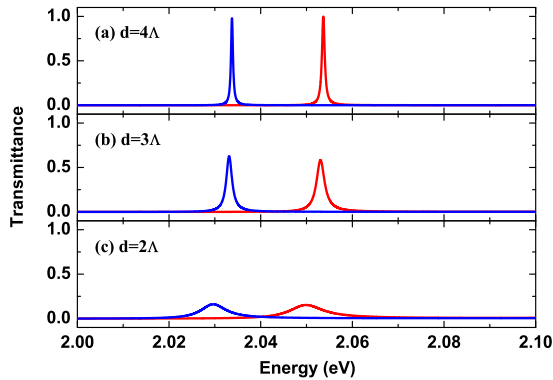}
\caption {\small (color online) Transmission spectra for a set of structures with different conducting overlayer thickness $d$ on a 12-period MPCs: (a) $d=4\Lambda$; (b) $d=3\Lambda$; (c) $d=2\Lambda$. The calculations are for $k_{x+}=0.3*2\pi/\Lambda$ (red line), and $k_{x-}=-0.3*2\pi/\Lambda$ (blue line), corresponding to the forward and backward illumination of light beams, respectively.}
\end{figure}

To support the above results, we have performed standard transfer matrix calculation of transmittance for the finite structure same as in Fig. 4. The results for the transmittance as a function of the conducting overlayer with finite thickness $d$ are shown in Fig. 4 under the light illumination of a plane wave. For larger conducting layer thickness of $d=4\Lambda$, almost full transmission is obtained along the forward direction at an energy near $E_{+}$, while complete reflection is obtained in the opposite direction at the same energy. Instead, the transmission peak along the backward direction appears at near $E_{-}$. Therefore strong nonreciprocity (given by $T(k_{x+})-T(k_{x-})$) at energies $E_{+}$ and $E_{-}$ is then achieved. These results are in excellent agreement with the infinite conducting overlayer dispersion curve results.  As the thickness of the conducting layer is reduced, the transmission maximum decreases, broaden, and shifts to lower energy, which is due to the insufficient ability to confine the surface state by thinner conducting layer. For smaller in-plane wave vectors and smaller angles of incidence (not shown here), thicker conducting layers are required in order to obtain narrow spectral features.

To take into account the loss effect in the conducting overlayer, we have carried out calculations in Fig. 5 for the cases of different values of the collision frequency, $\gamma$. As the loss is introduced, particularly in the case of high $\gamma$, transmission may become negligible and significant absorption can be observed in the structure. Fig. 5 shows the absorption and transmission spectra for a lossy conducting layer of thickness $d=2.2\Lambda$ for the in-plane wave vectors $k_{x+}$ and $k_{x-}$. As the collision frequency is increased from $\gamma=10^{12}$ rad s$^{-1}$ up to a maximum value $10^{14}$ rad s$^{-1}$, there is an initial trend to a higher but broader absorption maximum when the structure is forward illuminated by light beams (shown by the solid red line in Fig. 5). However, the situation becomes quite different in the opposite direction, and the absorption, also with transmission efficiency tends to be weaker, even at energies near $E_{-}$ (shown by solid and dotted blue lines in Fig. 5, respectively). That is to say, the consequence of the presence of such TPPs is visible in the absorption spectra, only when the structure is under front illumination. This can be clearly seen from the field patterns in Fig. 6. Fig. 6(a) shows that nearly full absorption can be achieved for front-illuminated structure, while the back-illuminated structure acts as a reflector shown in Fig. 6(b). Therefore, in the presence of the loss in conducting metal oxides, the proposed structure shows one-way absorption or reflection. Also, it can be seen that the higher loss can give rise to stronger and broader-band nonreciprocity in absorptance.

\begin{figure}[!htbp]
\centering
\includegraphics[width=3.2in]{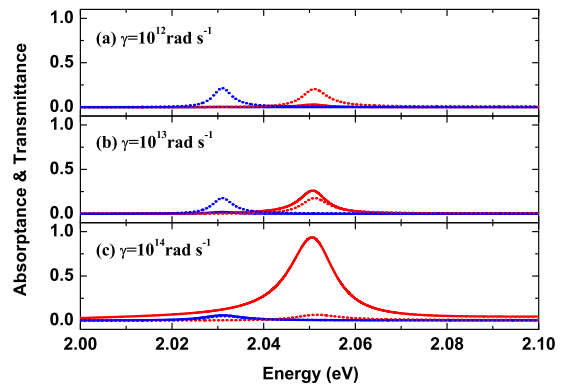}
\caption {\small (color online) Absorption (solid lines) and transmission (dotted lines) spectra for the structure with thickness $d=2.2\Lambda$ of conducting overlayer as a function of collision frequency: (a) $\gamma=10^{12}$ rad s$^{-1}$; (b) $\gamma=10^{13}$ rad s$^{-1}$; (c) $\gamma=10^{14}$ rad s$^{-1}$. Red and blue lines show the cases for $k_{x+}=0.3*2\pi/\Lambda$ (red line), and $k_{x-}=-0.3*2\pi/\Lambda$, respectively.}
\end{figure}

\begin{figure}[!htbp]
\centering
\includegraphics[width=3in]{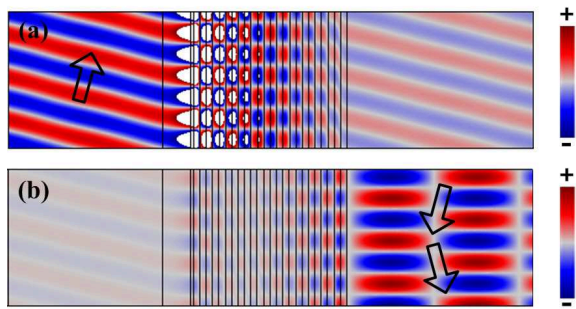}
\caption {\small (color online) Out-of-plane magnetic field patterns for the similar structure in Fig. 3, but the loss is introduced in conducting metal oxides, under front illumination (a) and back illumination (b). In this case, the excited TPPs energy $E_{+}$ shifted slightly to $2.051$ eV, the incident angle is also changed to be $75.93^\circ$, the collision frequency $\gamma=10^{14}$ rad s$^{-1}$, and the conducting layer thickness is assumed to be $d=2.2\Lambda$.}
\end{figure}

\section{\label{sec:level1}CONCLUSION}

In summary, we have studied that the nonreciprocal TPPs can be supported at the interface between the conducting metal oxides and MPCs at frequencies above the bulk plasma frequency. The key conditions require that the dispersion curve of TPPs falls into the photonic band gap of MPCs, and simultaneously lies outside the light cone for the conducting materials. The different positions of the peaks in the transmission spectra further support the presence of spectral splitting of TPPs for the cases of front and back illumination, which agrees well with the dispersion curves we have obtained. When the loss in conducting materials is considered, the behavior of nonreciprocal TPPs can still be observable in the absorption spectra. The results can be extended to more general systems provided that the required conditions are satisfied.

\section{\label{sec:level1}ACKNOWLEDGMENTS}

This work was supported in part by the National Science Foundation of China under Grant Nos. 11204036, 60990320, 60990324, 61138001, and 60921063, the 111 Project under Grant No. 111-2-05, and by the Campus Funding No. NY210050 from Nanjing University of Posts and Telecommunications. We thank Prof. Kin Hung Fung for useful discussions.


\begin{thebibliography}{99}
\bibitem{Tamm} I. E. Tamm, Phys. Z. Sowjetunion \textbf{1}, 733 (1932).
\bibitem{Yeh1} P. Yeh, A. Yariv, and C. S. Hong, J. Opt. Soc. Am. \textbf{67}, 423 (1977).
\bibitem{Yeh2} P. Yeh, A. Yariv, and A. Y. Cho, Appl. Phys. Lett. \textbf{32}, 104 (1978).
\bibitem{Vinogradov} A. P. Vinogradov, A. V. Dorofeenko, S. G. Erokhin, M. Inoue, A. A. Lisyansky, A. M. Merzlikin, and A. B. Granovsky, Phys. Rev. B \textbf{74}, 045128 (2006).
\bibitem{Goto} T. Goto, A. V. Dorofeenko, A. M. Merzlikin, A. V. Baryshev, A. P. Vinogradov, M. Inoue, A. A. Lisyansky, and A. B. Granovsky, Phys. Rev. Lett. \textbf{101}, 113902 (2008).	
\bibitem{Kawata} \emph{Near-Field Optics and Surface Plasmon Polaritons}, edited by S. Kawata (Springer, Berlin, 2001).
\bibitem{Kaliteevski} M. Kaliteevski, I. Iorsh, S. Brand, R. A. Abram, J. M. Chamberlain, A. V. Kavokin, and I. A. Shelykh, Phys. Rev. B \textbf{76}, 165415 (2007).
\bibitem{Shelykh} I. A. Shelykh, M. Kaliteevski, A. V. Kavokin, S. Brand, R. A. Abram, J. M. Chamberlain, and G. Malpuech, Phys. Status Solidi A \textbf{204}, 522 (2007).
\bibitem{Sasin} M. E. Sasin, R. P. Seisyan, M. A. Kalitteevski, S. Brand, R. A. Abram, J. M. Chamberlain, A. Yu. Egorov, A. P. Vasil'ev, V. S. Mikhrin, and A. V. Kavokin, Appl. Phys. Lett. \textbf{92}, 251112 (2008).
\bibitem{Gong} Y. Gong, X. Liu, H. Lu, L. Wang, and G. Wang, Opt. Express \textbf{19}, 18393 (2011).
\bibitem{Zhou} H. Zhou, G. Yang, K. Wang, H. Long, and P. Lu, Opt. Lett. \textbf{35}, 4112 (2010).
\bibitem{Zhang} W. L. Zhang and S. F. Yu, Opt. Commun. \textbf{283}, 2622 (2010).
\bibitem{Symonds} C. Symonds, A. Lemaitre, E. Homeyer, J. C. Plenet, and J. Bellessa, Appl. Phys. Lett. \textbf{95}, 151114 (2009).
\bibitem{Grossmann} C. Grossmann, C. Coulson, G. Christmann, I. Farrer, H. E. Beere, D. A. Ritchie, and J. J. Baumberg, Appl. Phys. Lett. \textbf{98}, 231105 (2011).
\bibitem{Gazzano} O. Gazzano, S. M. de Vasconcellos, K. Gauthron, C. Symonds, J. Bloch, P. Voisin, J. Bellessa, A. Lemaitre, and P. Senellart, Phys. Rev. Lett. \textbf{107}, 247402 (2011).
\bibitem{Kavokin} A. Kavokin, I. Shelykh, and G. Malpuech, Appl. Phys. Lett. \textbf{87}, 261105 (2005).
\bibitem{Liew} T. C. H. Liew, A. V. Kavokin, T. Ostatnick$\acute{y}$, M. Kaliteevski, I. A. Shelykh, and R. A. Abram, Phys. Rev. B \textbf{82}, 033302 (2010).
\bibitem{zongfu} Z. Yu, G. Veronis, Z. Wang, and S. Fan, Phys. Rev. Lett. \textbf{100}, 023902 (2008).
\bibitem{Brand} S. Brand, M. A. Kaliteevski, and R. A. Abram, Phys. Rev. B \textbf{79}, 085416 (2009).
\bibitem{Khanikaev} A. B. Khanikaev, A. V. Baryshev, M. Inoue, and Y. S. Kivshar, Appl. Phys. Lett. \textbf{95}, 011101 (2009).
\bibitem{Liscidini} M. Liscidini and J. E. Sipe, J. Opt. Soc. Am. B \textbf{26}, 279 (2009).
\bibitem{zongfu2} Z. Yu, Z. Wang, and S. Fan, Appl. Phys. Lett. \textbf{90},121133 (2007).

\end{thebibliography}
\end{document}